**Nova Southeastern University**

**NATIONAL INFRASTRUCTURE CONTINGENCIES:**
**Survey of Wireless Technology Support**

By

Ronald M. Fussell, M.S.

NATIONAL INFRASTRUCTURE CONTINGENCIES:
Survey of Wireless Technology Support

Ronald M. Fussell

*Nova Southeastern University*

In modern society, the flow of information has become the lifeblood of commerce and social interaction. This movement of data supports most aspects of the United States economy in particular, as well as, serving as the vehicle upon which governmental agencies react to social conditions. In addition, it is understood that the continuance of efficient and reliable data communications during times of national or regional disaster remains a priority in the United States. The coordination of emergency response and area revitalization / rehabilitation efforts between local, state, and federal emergency response is increasingly necessary as agencies strive to work more seamlessly between the affected organizations. Additionally, international support is often made available to react to such adverse conditions as wildfire suppression scenarios and therefore require the efficient management of workforce and associated logistics support.

Various governmental and civilian organizations have been formed to serve as intermediaries for data communications during disaster response scenarios in an effort to streamline command and control channels for information. The preeminent disaster response agency in the United States federal government, the Federal Emergency Management Agency (FEMA), has filed the "Fiscal Year, 2000 Disaster Activity Overview" report on major national disasters resulting primarily from extreme weather conditions and geographic occurrences [FEMA, 2000]. In addition, the need for wildfire suppression in various states resulted in declarations of emergency throughout the reported year.

Various methods of supporting data transactions are available for use during nominal life activities in most communities such as terrestrial and extra-terrestrial based communication channels. However, during most emergencies, terrestrial connectivity is lost or diminished as a result of the occurrence and on occasion, the act of recovery from the occurrence. Therefore, it is necessary to examine the technological horizon to discover those technologies best suited to support during times of infrastructure contingencies. In addition to the examination of available technologies for use during emergency response and recovery operations, an investigation into methods of ensuring the integrity of transmitted data between parties must be undertaken.

As the reliance upon wireless and other extra-terrestrial means of data communication increases as a result of disaster conditions, the accessibility of unauthorized persons to transmitted data increases as the data uses the environment as the transfer medium. The susceptibility of the data to interference and unauthorized monitoring causes concern from agencies managing such needed response and recovery actions as law enforcement and security, health and welfare, infrastructure recovery, and family services. Recent events have caused an increased sensitivity of the general United States populace to issues of homeland security. This sensitivity extends beyond the personal safety and security to encompass most environmental interfaces, including the interactivity between users of transmitted data.

It is through the examination of the issues related to un-tethered data transmission during infrastructure contingencies that responders may best tailor a unified approach to the rapid recovery after disasters occur.

**Introduction**

The need for reliable information continues to drive modern research in many directions. Research continues to focus upon new methods of transporting data across a variety of medium in an effort to accelerate the data arrival to almost real-time. Through this capability the potential for remote commanding and monitoring of critical equipment and systems increases in both scope and applications. Disaster Response Plans also become re-engineered to capitalize upon the capabilities found in the employment of innovative concepts, such as distributed teaming and remote commanded operations. Through the use of wireless and satellite data distributed systems, remote data collection and distribution to affect the distribution of resources, assessment of critical situations, as well as, notifications of degrading emergency conditions and potential personnel hazards are capable.

This survey applies the technology currently available to support ubiquitous operations to the operational view of disaster response. In this survey, certain operational assumptions are understood and integrated into the document scope.

Initially, it must be cited that the document addresses instances of catastrophic failure in the national terrestrial infrastructure for data communications. This failure could reasonably be as a result of natural occurrence, an event of terrorism, or a result of civil activities beyond that planned as nominal occurrence.

Secondly, although many of the aspects of the examined technologies are applicable to most recovery operations, the examination of current technologies used by the United States military, intelligence communities, and other specialized governmental agencies are specifically not covered in this survey. This is due to the potential for disclosure of technologies presently held in confidence where disclosure could jeopardize the capabilities of that agency to effectively accomplish the assigned mission. Additionally, it must be cited that this research centers primarily upon data transfer technologies that utilize a radio frequency emission as the carrier for the transmitted information. This scope is primarily due to the general issues related to other current technologies that support ubiquitous computing and data transmission. Technologies such as infrared (IR), ultrasonics, and directed light energy service environments that are often less dynamic than the emergency response operation primarily due to line-of-sight, range, energy reflection and refraction, and issues regarding general transmission integrity. Therefore, these technologies are not within the scope of this survey.

Finally, this survey recognizes the differing levels of incidents, operational requirements, and differing levels of data security in response to the event under response conditions. However, this survey addresses the issues regarding security through the discovery and relating of data processes and does not attempt to define specific technologies or algorithms for employment in emergency response conditions. Therefore, it is incumbent upon emergency responders to develop a comprehensive plan prior to need.

**Research Scope**

In order to achieve an understanding of the potential data transfer technologies that may be usable during disaster response and recovery, an examination of the government al stakeholders in the response scenario. Through the identification of the general notification process of responders to a typical catastrophic event, available resources are more clearly defined. Therefore, a brief familiarization of the governmental agencies involved is undertaken in this document.

This research is founded upon the premise that during most contingencies of infrastructure, the normal channels of data flow are disrupted. During these periods, reliance upon mobile, autonomous collection and distribution of data is required to support recovery operations. This survey assumes only to identify the technologies used during the contingencies to better support recovery operations and excludes contingency plans related to commercial endeavors.

Furthermore, this research is focused upon technologies and techniques predominate to the use of radio frequency and microwave transmissions to support reliable, high volume data traffic.

In addition, it must be noted that this research is undertaken with the support of emergency responders to natural and civil disasters in mind. Therefore, the support of responders to threats of terrorism or attacks upon United States infrastructure elements by agents of foreign entities are not specifically addressed in this research. However, it is due to the similarities in general recovery actions required to restore nominal data distribution for a locale that many of the technologies and techniques may be used in either case of terrorism or natural disaster. Also, many of the channels for coordination and support may be similar under the recently created U.S. Office of Homeland Security to support the response to and recovery from such attacks.

**Historical Overview**

Historically, the transfer of data during disaster situations was accomplished through a network of volunteer amateur radio operators that supported the transfer of data messages for various unclassified military origins and civil organizations. The first attempt by the federal government to offer national recognition and organizational support to the respondents to disaster occurrences that provided communications services occurred in 1952. At that time, the organization, known as the Radio Amateur Civil Emergency Service (RACES) organization was conceived. Currently, this organization continues to support during times of contingency for communications and is staffed by a volunteer communications group within government agencies. Originally concerned with voice traffic during contingent times, the radio environment has grown to accommodate the transfer of packet data for emergency response organizations. With added services as support to search and rescue operations, signal location / triangulation, and messaging support, the organization continues to render localized and regional support of emergency activities.

Presently, the organization tasked with the management and coordination of communication services to support contingency efforts on a national scale is the National Communication System, (NCS).

The origins of the NCS began in 1962 after the Cuban missile crisis when communications problems among the US, the USSR, NATO, and foreign heads of state threatened to complicate the diplomatic mission. President John F. Kennedy ordered an investigation of the national security communications causing the National Security Council (NSC) to conduct a committee-led investigation of the capabilities to support critical communications services. The results included a recommendation to the President to charter an organization to oversee such critical functions. Consequently, in order to provide better communications support to critical United States Government functions during emergencies, President Kennedy established the National Communications System (NCS) by a Presidential Memorandum on August 21, 1963. The NCS mandate included linking, improving, and extending the communications facilities and components of various Federal agencies, focusing on interconnectivity and survivability.

It was on April 3, 1984 that President Ronald Reagan signed Executive Order 12472 which broadened the scope of NCS involvement in matters of national security and emergency preparedness capabilities. This re-scope also resulted in the involvement of twenty-three federal agencies in roles ranging from resource suppliers to resource subscribers. This involvement allows for the timely response to service requests by the NCS organization.

The National Communications System (NCS) has undertaken a number of initiatives to provide communications to support hazardous scenarios of varying degrees. One such program is the Shared Resources (SHARES) High Frequency (HF) Radio Program.

The purpose of the SHARES program is to provide a single organization to support the transmission and distribution of message data using existing High Frequency radio resources of Federal, state and industry organizations when normal communications are unavailable. This service is provided to support the transmission of national security and emergency preparedness information.

The SHARES network consists of radio stations representing Federal, state, and industry resource contributors and are located across the continental United States and specific locations overseas.

The use of messaging via high frequency radio equipment remains a viable means of data transmission in the current technological scheme. Local and regional Emergency Management Services are encouraged by the federal government to utilize volunteer amateur radio organizations to act as the first line of response in most disaster scenarios.

**Theory and Research**

Disaster response and recovery operations are undertaken based upon the types of emergency encountered. Traditional first response to emergency conditions is founded upon local guidelines that are specific to the geographic area under the emergency conditions. As the contingency event goes larger in scope or requires more specialized support, the local Emergency Management or Civil Defense coordinators will contact the next level of responder in order. The general order of response is cited below.

- ➢ Local Emergency Management System
- ➢ County or Providence Emergency Management System
- ➢ State Emergency Management System
- ➢ Regional Emergency Management System
  (This coordination is generally undertaken through the Federal Emergency Management Agency).
- ➢ National Emergency Management System (Federal Emergency Management Agency).

Historically, the role of communications during infrastructure contingencies has been primarily filled by means of voice or signal communications. This level of communications had previously been used to relay emergency instructions and requests for support of personnel in the disaster response role by means of messaging. In more recent history, the means of transferring data via a wireless medium has increased the capabilities of disaster recovery management to coordinate efforts, especially in the arena of logistics support and real-time operations. In order to best understand the technologies that may be used to support such data-related contingencies, one must understand the scope of the disasters that threaten the nominal flow of data in the United States infrastructure.

**Natural Disasters**

Examination of typical disasters that threaten infrastructure supportability of nominal data traffic is undertaken by a review of a typical year in disaster response. A review of the Federal Emergency Management Agency (FEMA) document, "Fiscal Year 2000, Disaster Activity Overview", published on December 1, 2000, best describes the typical large-scale threats to infrastructure and thereby assumes a growth of the threat from localized to the larger scale.

During the year, according to the FEMA document, disasters ranging from the Year 2000 bug (issues associated with the Y2K arrival) to wildfires and weather related issues. Although the table below lists differing incidents, only incidents reported to the FEMA response teams are cited.

**FY 2000 Reported Disasters**

| Incident Type | Total Declarations | Affected States |
|---|---|---|
| Severe Storms (some including flooding or tornadoes) | 19 | AZ, KY, GA, WV, OH, AL, TX, ME, KS, MO, TN, WI, MN, ND, NY, NJ, DC |
| Severe Winter | 9 | LA, AL, GA, SC, NC, |
| Storm | | VA, DC, MD, SD |
| Tropical Weather System | 5 | USVI, FL, VT, NH, ME |
| Wildfires | 3 | NM, MT, ID |
| Flash Flooding | 1 | VT |
| Avalanche | 1 | AK |
| Earthquake | 1 | CA |

[FEMA, 2000]

**Disaster Planning**

Disaster plans are most often based upon simulations, exercise, or modeling outcomes and represent the information developed from mathematical or physical representations of a process, entity, phenomenon, or system, as cited in the Information Technology Research for Crisis Management [Computer Science and Telecommunications Board, 1999]. This research notes the major roles of the modeling and simulation process to include the identification of areas of specific concern during a contingency event and alternate methods of supporting recovery efforts. Such areas as data traffic routing, system prioritization, and alternate data transfer methods are integral to the planning of disaster recovery. Additionally, the mitigation of risk in operations during an incident may be accomplished through the education of response teams prior to the occurrence of such an event, the identification of resource conflicts and possible resolution, and the establishment of command and control channels used to facilitate the contingency efforts.

In keeping with the spirit of the United States Presidential Decision Directive 63 and other relevant United States Statutes, systems utilized to support contingency operations will reflect the most current elements of system and data security. A security survey and associated plan is performed per relevant directives and criteria in an effort to characterize system and data integrity threats prior to need and incorporate these system requirements into the Disaster Response Plan. Guidance in the system survey derives technical credibility from sources such as the International Common Criteria (CCITSE), Trusted Computer System Evaluation Criteria (5200.28-STD, Orange Book, as revised) [National Security Agency], and relevant National Communications System directives. Primary areas of focus in the survey include those listed below.

- Password Management
- Audit Functions
- Discretionary Access Control
- Trusted Network Interpretation
- Configuration Management
- Identification and Authentication
- Object Reuse and Covert Channels (As appropriate)

The results of the security survey act as the focus in the planning for security process, procedures, and technology developments / procurements / implementations. The planning of system and data security considers the distributed nature of the network and associated test team access in an effort to assure protection from threats of domestic and foreign origination. Security structure is founded upon the categorization of threats and the asset to be protected. Such structure includes protection for operating system elements, database management systems / related data, network assets, and Internet specific interfaces.

**Wireless Support**

Wireless networks offer many solutions during contingency response efforts due to the potential for autonomous operation, frequency management schemes, and portability. Historically, wireless data subscribers viewed as a rapidly emerging technical sector of data users. In 1996, the Computer Science and Telecommunications Board published "The Unpredictable Certainty" in which was cited statistical growth of wireless users from 500,000 user in 1995 to a forecasted 9.9 million in the year 2000 [Computer Science and Telecommunications Board, 1996]. Thereby indicating the perceived capabilities of the technology to support a wide variety of data types and applications.

It is this capability that is realized during the response to disaster.

Technologies, such as Global System for Mobile Communications (GSM), Bluetooth networking initiatives, and Wireless Applications Protocol promise the user of the ubiquitous network with the mobility previously unattainable from legacy networks.

A GSM network is comprised of several functional entities, whose functions and interfaces are specified and standardized (GSM Alliance, 2000). The GSM network can be divided into three broad functional elements. The subscriber of the service obtains a mobile station that interfaces with the public GSM networking infrastructure to provide the cybernetic link. The duplex radio frequency transmissions with the Base Station Subsystem are controlled from within that provider's base station. The Network Subsystem, the main part of which is the Mobile Services Switching Center (MSC), performs the switching of calls between the mobile users, and between mobile and fixed network users. The MSC also handles the mobility management (handoff and frequency control functions) operations.

Designed to operate in a noisy radio frequency environment, the Bluetooth radio uses a fast acknowledgement and frequency-hopping scheme to make the link robust. Developed through the cooperation of 3Com, Ericsson, Intel, IBM, Lucent, Microsoft, Motorola, Nokia, and Toshiba, Bluetooth radio modules avoid interference from other signals by hopping to a new frequency after transmitting or receiving a packet. Bluetooth technology allows for the replacement of the many proprietary cables that connect one device to another with one universal short-range radio link. For instance, Bluetooth radio technology built into both the cellular telephone and the laptop would replace the cumbersome cable used today to connect a laptop to a cellular telephone. (Motorola, 2000).

While GSM and Bluetooth promise to open mobility boundaries for the user, Wireless Applications Protocol (WAP) will provide the means of exploiting the un-tethered platform. WAP bridges the gap between the mobile world and the Internet, as well as, corporate intranets and offers the ability to deliver an unlimited range of mobile value-added services to subscribers—independent of their network, bearer, and terminal. Mobile subscribers can access the same wealth of information from a pocket-sized device as they can from the desktop.

For wireless network operators, WAP promises to decrease operations turnover while cutting operational costs and increase the subscriber base both by improving existing services, such as interfaces to voice mail, facilitating an range of new value-added services and applications, such as account management and support tracking. New applications can be introduced quickly and easily without the need for additional infrastructure or modifications to the phone.

Applications are written in wireless markup language (WML), which is a subset of extensible markup language (XML), a widely used interfacing language. Using the same basic model as the Internet, WAP will enable content and application developers to grasp the tag-based WML that will pave the way for services to be written and deployed within an operator's network quickly and easily. As WAP is a global and interoperable open standard, content providers have immediate access to a wealth of potential customers who will seek such applications to enhance the service offerings given to their own existing and potential subscriber base.

WAP-compliant phone uses the in-built micro-browser to make a request in WML (Wireless Markup Language), a language derived from HTML especially for wireless network characteristics. This request is passed to a WAP Gateway that then retrieves the information from an Internet server either in standard HTML format or preferably directly prepared for wireless terminals using WML. If the content being retrieved is in HTML format, a filter in the WAP Gateway may try to translate it into WML. A WML scripting language is available to format data such as calendar entries and electronic business cards for direct incorporation into the client device. The requested information is then sent from the WAP Gateway to the WAP client, using whatever mobile network bearer service is available and most appropriate (IEC, 2000).

Initially, services are expected to run over the well-established SMS bearer, which will dictate the nature and speed of early applications. Indeed, GSM currently does not offer the data rates that would allow mobile multimedia and Web browsing. With the advent of general packet radio services (GPRS), which aimed at increasing the data rate to 115 kbps, as well as other emerging high-bandwidth bearers, the reality of access speeds equivalent or higher to that of a fixed-line scenario become evermore believable (African Cellular, 1999). General packet radio services technology, is viewed by many technologists as a fitting partner for WAP, with its distinct time slots serving to manage data packets in a way that prevents users from being penalized for holding standard circuit-switched connections,

**Wireless Issues**

Initially, the method of data transfer in the wireless domain must be examined in order to determine characteristics and ultimately, resolution of technical issues. When investigating wireless data networking operations, examination of the operational environment will reveal most technical issues. A common assumption among researchers in this venue is the "Concept of Free Space" as outlined in Digital Communications: Fundamentals and Applications, second edition. In this concept it is assumed that the radio frequency channel is free of hindrances to signal propagation. Such impediments as moisture and materials that may cause absorption, reflection, refraction or diffraction may cause the transfer of information to become unreliable when using radio frequency transmission techniques. These impediments may be manifested as losses due to signal-to-noise ratio or signal distortion due to inter-symbol interference (Sklar, 2001). A general discussion of wireless transmission impediments follow, however, inter-symbol interference discussion is beyond the scope of this survey.

Environmental conditions play an important part in the selection of the wireless platform to support emergency response operations. Such common concerns as the presence of hazardous commodities (potential of interactivity or explosion), electromechanical or radio frequency interference sources, and air of high moisture content may offer operational impediments or impede the reliability of data transmissions using radio-frequency emissions. Therefore, it must be cited that the wireless platform is not considered a "wholesale option" for use even though providing a wide range of capability and applications to support the operational aspects of disaster recovery.

Issues related to the range of transmission and system data rates also become and issue of wireless operations. While the wireless platform offers a variety of techniques to support data transfer, such as seamless hand-off between cells (or micro-cell sites) and peer-to-peer communications, data relay, and storage, the statistical potential for data packet loss increases with each hand-off of the information. Additionally, it must be assumed that to adequately utilize the wireless platform, a cell must be available in the proximity of the recovery operations. This requires either the connectivity of that cell to functioning infrastructure elements or satellite communications capability.

The integrity of data becomes an issue of paramount concern during critical operations such as those associated with the emergency response and recovery following a catastrophic event. The number of attacks upon governmental and commercial repositories of data has escalated in recent years. It is reported that economic losses associated with reported security breaches for the first quarter of the year 2000 that through theft alone were in excess of $66 million dollars [Power, 2000]. Additional losses have been also realized from cyber-attacks including financial fraud, virus, sabotage, and insider net abuse and indicate the magnitude of losses to the current economy as a result of dubious or missing data.

With the need for data integrity, it is required that an understanding of the relief available for the management of the data communications activity. Most of the active malicious system attacks fall under the auspices of the United States Federal Criminal Code, specifically, Title 18, Section

Chapter 1030 (Computer Fraud and Misuse Act) and Section Chapter 90 (Economic Espionage Act) guidelines. The Computer Fraud and Misuse Act was passed and signed in 1986 provides federal criminal guidelines for prosecuting persons who intentionally access a computer of which one is not authorized or exceeding established authorization in order to retrieve information of which one is not entitled. This includes government computers and computers used in interstate or foreign trade related transactions. The Economic Espionage Act was passed in 1996 and precludes one from profiting from the misappropriation of another's trade secret by making it a federal offense through the language regarding "upload", "download" and "e-mail" (United States Federal Criminal Code, 2000).

Although the protection of data under the law is an important aspect of the present socio-economic environment, it must be noted that the laws have predominately addressed the data flow using terrestrial capabilities. Therefore, these technologies predominately apply to tethered network architectures, other terrestrial network assets, and data in general. The preference for hard-line access of information is often transcended by necessity during periods of emergency or infrastructure contingencies by mobile, wireless and space access data communications. Gone are the days of immobile organizations and the concept of centralized information access. In today's society, the flow of information is from the site of any emergency to the supporting agencies directly, which enables the user to employ the most current data in the management of the decision making process to support contingent efforts. The availability of current information in remote geographical areas is crucial to the successful and timely completion of tasks. It is through flexible connectivity that the proliferation of data relating to technical topics, as well as, resource support tasks may best be disseminated throughout the affected community. Therefore, the need for security of transmitted data becomes increasingly important as the terrestrial infrastructure becomes more sophisticated, as is the means of attack upon the data medium.

Additionally, the security and integrity of data in the network has become of paramount concern following the identification by governmental and industry sources of various threats to the nominal transfer of data through common infrastructure components. Unauthorized replication, modification, and other fraudulent use of network resources cost industry and government agencies resources more with each passing year [Allen, 2001]. Statistics reveal that economic losses associated with reported security breaches for the first quarter of the year 2000 that through theft alone were in excess of $66 million dollars [Power, 2000]. Other methods of loss include financial fraud, virus, sabotage, and insider net abuse and represent a rising trend in annual losses and indicate the magnitude of losses to the current economy as a result of dubious or missing data. The table below delineates the current financial losses for the year, 2000, to date as a result of dubious data as reported by the Computer Security Institute for the United States Federal Bureau of Investigation [FBI/CSI, 2000].

**Year 2000 Security Losses**

| Types of Security Breaches | Value of Loss |
|---|---|
| Theft of proprietary data | $66,708,000 |
| Unauthorized insider access | $22,554,500 |
| Telecom fraud | $4,028,000 |
| Financial fraud | $55,996,000 |
| Virus | $29,171,700 |
| Laptop theft | $10,404,300 |
| Insider network abuse | $27,984,740 |
| Denial of service | $8,247,500 |
| Sabotage | $27,148,000 |
| System penetration | $7,104,000 |
| Telecom eavesdropping | $991,200 |
| Active wiretapping | $5,000,000 |

[FBI/CSI, 2000]

In an effort to understand the magnitude of the security issue, one must determine the scope of the term, security. An initial view leads one to examine the survivability of a network or related assets to an attack. This measure holds an important key to the integrity of the data resident on that system and alludes to the flexibility of network assets to cope with internal, as well as external intrusions or misuse [Allen, 2001]. Furthermore, the survivability of a network and the associated economics must be assured regardless of the transmission or storage media. However, the economics associated with security issues is not limited to the system level. The focus of most security issues relates to the integrity and usability of the data resident within the confines of the identified system and is generally thought to infer a potential data of destruction.

Technologies available to support data security initiatives include, but are not limited to, the encryption of data thereby affording the information unintelligible and using transmission techniques that render the transmitted signal beyond reception. Encryption relies upon the alteration of the information by mathematical algorithm in order to mask key features of the transmitted data thereby rendering the data unintelligible to those monitoring the data without the proper decryption algorithm.

There are two general forms of encryption of transmitted data each using a lock and key scenario. The first employs a 'secret key' and is normally used for point-to-point transactions with previously designated users. The second form of encryption uses a private key and a public key scenario to ensure authentication. This algorithm eliminates the need for common trust between transmitter and receiver. This is accomplished by employing the use of the product of two very large prime numbers (>10 to the $100^{th}$ power). This protocol relies upon the notion that determination of products of such large numbers is computationally intensive and economically unfeasible to decipher. These methods of encryption are used predominantly as a method of securing the data prior to transmission [Stallings,1999].

Transmission techniques that render the transmitted signal beyond reception is often accomplished in the form of variations of the transmit frequency. Techniques, such as direct sequence spread spectrum or frequency hopping spread spectrum, create variations of the transmitted signal thereby decreasing the probability of unauthorized monitoring. Frequency Hopping Spread Spectrum transmissions modulates the outbound data with a carrier signal that hops from frequency to frequency as a function of time and based upon a programmed algorithm. Direct Sequence Spread Spectrum combines a data signal at the transmission source with a greater data rate bit sequence that is often referred to as a chipping code. Through this combination, the number of received data bits is often ten to twenty times more than the original data thereby creating a bit stream where the actual data is masked.

**Satellite Support**

Readers of the October edition of Wireless World in 1945 became some of the first to witness a vision of the space borne wireless relay station in the acclaimed article, "Extra-Terrestrial Relays" by Arthur C. Clarke. It was through this vision that the contemporary communications satellite has evolved into a vehicle to further the reach of society into geographic regions and across barriers where the construction of a communications infrastructure is prohibitive, thereby providing the basis for unlimited communications mobility.

Satellite communications has become the cornerstone of the current telecommunications in times of disaster or infrastructure failure. For example, due to the magnitude of the devastation caused by Hurricane Andrew, the Federal Emergency Management Agency and the Joint Task Forces were required to work together to accomplish the mission of providing support of federal agencies in the rescue of persons, securing of property, clean-up tasks, and interim support prior to infrastructure repair. During this emergency effort, cellular telephones were ineffective as a reported ninety percent blockage rate of calls was cataloged by the Federal Emergency Management Agency. Therefore it was necessary to resort to alternative traffic support measures from terrestrial microwave to satellite line to support the contact with outside support sources. Although during this effort, no attempts to gain access to the transmitted data were cited by the Federal Emergency Management Agency, it remains a concern as the nature of infrastructure contingencies evolve. From natural disasters to malicious attacks of cyber-terrorists, the scope of system security must consistently guard against such acts as listed below [United States Army, 1997].

- Unauthorized access
- Theft of proprietary data
- Data sabotage or disruption
- Creation or distribution of malicious code

In one example of the need to use extraterrestrial data relay comes from an attack of terrorism in which a method of secure communications was needed to support the subsequent investigation. One month prior to the 1996 Olympics, the Department of Justice requested communications support from the National Communication Service in case of emergencies during the games. The National Communication Service, through coordination with the NCS member, National Aeronautics and Space Administration, identified the Advanced Communications Technologies Satellite and the associated ACTS Mobile Terminal (AMT) with several American Mobile Satellite Corporation (AMSC) briefcase terminals as the solution to this request. The equipment was subsequently transferred to the Atlanta area and staged to support emergency needs. The AMSC terminals were provided to the U.S. Marshals Service and the AMT was to remain in standby mode until the Federal Bureau of Investigation required the support.

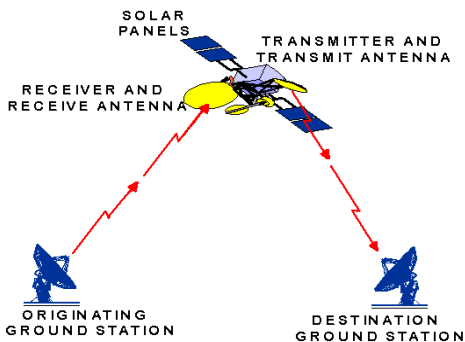

Uplink, Downlink, and Satellite Component Illustration
[NASA ACTS program, 1998]

At about 0345 on 27 July of that year, the Federal Bureau of Investigation contacted the appropriate parties to move the terminals to a remote site at Centennial Olympic Park and was located next to a Federal Bureau of Investigation van, approximately 100 feet from the site of a bombing. Full communications from the ACTS Mobile Terminal to the satellite to the public switched network were made within one hour of the Federal Bureau of Investigation request. The ACTS Mobile Terminal then provided ten 64Kbps lines into the Federal Bureau of Investigation van and two 64Kbps lines to an outdoor tent. The end user equipment consisted of standard phones, cordless phones, and fax machines provided by both National Aeronautics and Space Administration Jet Propulsion Laboratory and the Federal Bureau of Investigation. Operations proceeded until 1900 hours, (7 o'clock p.m.), 27 Jul 96, when the Federal Bureau of Investigation communications was no longer needed in the area. Users of the capabilities included Federal Bureau of Investigation, Bureau of Alcohol, Tobacco, and Firearms, U.S. Marshal Service, and military/guard personnel [National Communications System, 2000].

**Satellite System Issues**

It is hypothesized that although a select amount of data may be encrypted prior to transmission, the key method for the prevention of interception and subsequent intelligence disclosure of satellite linked data streams may be found in the transmission medium and the manipulation of signals to achieve an efficient data transfer. The key to assuring the integrity of signals is, essentially, to place the intelligence in locations within the signal and in unanticipated formats that would assure security. Additionally, the economics associated with the interception, decoding and decryption of the information would be prohibitive for the general populace and thereby offering a measure of security in itself. As increased encryption of a signal, higher signal strength or frequency, signal directionality, or remote locations become involved in the transfer of intelligence, the more costly it becomes to breach the data stream for perpetrators of information sabotage or reception.

The devastation caused by Hurricane Andrew in August of 1992, the bombing of the Atlanta Centennial Olympic Park on July 27, 1996, and most recently, the attack upon the World Trade Center and Pentagon by Islamic extremists have resulted in situations where the use of the public communications or cellular communications are not feasible to support the investigation and recovery in the aftermath. In each of these situations, the use of satellite based communications relay has proven itself as a formidable tool in the management of information.

Although the satellite technology has evolved to become "switchboards in the sky", a number of issues remain that must be addressed prior to employment in a disaster recovery scenario. Initially, one must be concerned with the potential for personnel injury as most satellite systems that support burst transmission modes or high volume, long duration transmissions, operate in the microwave frequency spectrum and with relative high power. Additionally, care should be taken when operating such equipment in areas that may contain hazardous commodities due to the potential for ignition or explosion. Other issues include the availability of satellite transponders to support the relay and subsequent spot-beam downlink to a ground station for connection to functional national infrastructure elements. Satellite services such as INMARSAT, Intelsat, PanAmSat, DomSat, and government owned / operated vehicles, MilStar and Tracking and Data Relay Satellite (TDRS) provide relay support by allocation of transponders reserved on each craft. The location of the spacecraft is also important in the selection of the satellite link to facilitate the data requirements of disaster recovery.

Microwave systems are generally employed to provide communications linkage between ground sites and the orbiting vehicle. Microwave transmissions are essentially radio frequency radiation in the microwave spectrum (high megahertz range into the low gigahertz bands) that offers high bandwidth capacity with short wavelength operation [Littman, 1998]]Office of Spectrum Management, 1996]. Modulation techniques include the shifting of phase, frequency or amplitude to incorporate intelligence into the radiated signals. Also mentioned previously, satellite microwave transmissions often support long duration and burst mode transfers of data very well, however, it does carry certain restrictions in frequency allocation and operating considerations. One example is the Ku band transmission is perhaps most notably efficient for high density, long duration transactions and supports compression techniques well. Ku band transmissions support data streaming, and may be polarized for immunity to interference.

A reference to the previous discussion regarding the Concept of Free Space is necessary at this point. It must be noted that the disaster recovery operation may be underway as a result of unfavorable weather or other event that may cause atmospheric interference of the microwave signals and thereby preclude reliable data transfer by satellite link. Particularly, precipitation holds moisture particles in such density that it often correlates with the transmitted bandwidth. Other potential sources include geological or topographical elements such as iron ore and certain non-porous rock formations that may impede a line-of-sight signal [Glover, 2000]. Propagation delays during the transfer of data via satellite vary according to satellite orbit, location (apogee or perigee), and method of inter-satellite communications, if used. Potential users may calculate propagation delays using the following example as a reference.

---

Geosynchronous Orbit = approx. 22,300 miles altitude X 2 (uplink, downlink) = 44,600 miles
Calculated speed of light / radio emissions = 186,000 miles per second

44,600 miles signal travel/186,000 miles per second = .2398 seconds (239.8 milliseconds propagation delay)

---

[Stallings, 1997]

Due to the robust characteristics of the satellite system, various data types may be transferred using this method. The digital media capability exists which properly utilizes the broad bandwidth characteristics of the satellite system to transport audio, video, as well as, data. Security of the data takes place in differing methods depending upon the data type, satellite system architecture, and potential use. All encryption systems have three main components: the encoder, transmission media, and a network of decryption units assigned to specific users.

Video capture during the response to a catastrophic event may be used to characterize the event, as well as, plan remediation. Therefore, the need to encrypt the video transmission to assure integrity becomes evident. Video encryption is usually undertaken in less-evolved systems via analog methods. The analog systems most often encrypt the video by stripping away the vertical and horizontal synchronization pulses prior to the uplink of the signal. Without these control signals, conventional television receivers are unable to determine the initial tracing of each new scan line within a field or the beginning of the new field. The satellite uplink's encoder also creates a negative image by inverting the video signal. Additionally, certain analog systems relocate the chroma (color burst) signal to a non-standard frequency that is detectable by the integrated receiver-decoders intended. A special digital synchronization pattern is transmitted by the uplink commands for the integrated receiver-decoders to regenerate the proper horizontal and vertical pulses internally. In some encryption systems, Scene Change Detectors are used to detect scene changes and alternate the security scheme during intervals when the video content changes abruptly. Multiple, programmed/random, inversion modes are used to invert the applied video on either a field-by-field or individual line-by-line basis as an added measure of security. In the case of a system using a security smart card programmed with a specified code that is coincident to a synchronization scheme for the video, the integrated receiver-decoder is required to have a card that contains the coincident algorithm.

Currently, digital transmission is afforded in much the same way by digitizing the video and format information prior to transmission. Through the digitizing operation additional securing may be undertaken while simultaneously compressing the video information. Currently, two methods of channel coding are most prevalent in the coding of satellite transmission intelligence. The first is waveform coding and provides for improved detection and combats signal-fading effects. The second method of preventing noise and signal-fading effects while resisting signal jamming attempts is structured sequences. Structured sequences provide for data integrity through the use of error correction techniques based upon checksums and other indicators. One method requires the receiver to request for re-transmission from the transmitter via a series of automatic repeat request (ARQ) bits found in the uplink bit stream. Another technique uses information bit configuration to act as a reflection of the data that is transmitted.

Algorithms used during transmission include linear block coding, (including Reed-Solomon codes), convolutional, and the turbo coding concepts. The fundamental hardware unit for a convolutional encoder is a tapped delay line or a shift register with L+1 stages. Each tap gain (g) is a binary digit representing a short-circuit connection or an open circuit. The message bits in the register are combined by modulo-2 addition to form the encoded bit:

To provide the extra bits needed for error control, connecting two or more modulo-2 summations to the register and interleaving the encoded bits via a switch achieve an output rate greater than the message bit rate. In this situation, each message bit influences a span of n(L+1) successive output bits. The quantity n(L+1) is called the constraint length measured in terms of encoded output bits, whereas L is the encoder's memory measured in terms of input message bits. (Zhang, 2001). Transmission of the video stream in analog is via a high-powered microwave link to the satellite that may be polarized as dictated by atmospheric conditions. For example, left or right, circular polarization is often used to deter the effects of smoke, haze and precipitation. Polarization may take place either mechanically in the transmitting antenna through the use of a quarter-wave dielectric plate placed in a 45 degree angle from the transmitted energy phasing or electrically through magnetically positioning the transmitted wave in 45 degrees from the reference transmitted pulse or burst.

For reception, the user's integrated receiver-decoders address number is forwarded on to the Authorization Access Center. A specific authorization message is sent over the satellite activating the individual integrated receiver-decoders with instructions to decode the particular data feed in a specified format. For most governmental access channels, the Authorization Access Center is not required due to pre-arrangements with the servicing organization, such as the Goddard Space Flight Center (National Aeronautics and Space Administration). Most encryption systems employ tiers: special encrypted data codes that are allotted to using agencies. Each video provider is assigned one or more unique tier bits, with each bit capable of authorizing reception of either a single event or full-time access. The Conditional Access data is inserted into the vertical blanking interval of the video signal. The more lines allocated to this data stream, the greater the number of decoders that can be addressed in a given authorization period. Decryption may be through use of the Viterbi decoding algorithm or a complimentary technology such as feedback or sequential algorithms for convolutional encryption schemes. The Viterbi algorithm may be summarized formally as:

For each $i$, $i = 1, \ldots, n$, let:

$$\mathbf{X}_i = (X_{i_1}, X_{i_2}, \ldots, X_{i_r})$$

This initializes the probability calculations by taking the product of the initial hidden state probabilities with the associated observation probabilities.

The satellite system must possess the capability to communicate with other operationally associated systems and systems interfaces. Due to the critical nature of many emergency response operations, the system used to support the command and control of field operations as well as, offer remote commanding and control of support equipment as required. Command provisions are made to ensure that the system is capable of command and control functions during integrated operations as well. Management of data decommutation rates for Pulse Code Modulation or appropriate formatted telemetry at acceptable rates for the services required by emergency management and operations of remote sensors or support system control.

Digital encryption systems that service video and audio simultaneously convert the sound portion of each television program from an analog to a digital signal expressed as binary numbers which correspond to the 'off' (0) and on (1) logic states of computer circuits at the uplink by the audio encoder. Generally, analog audio signals are transmitted in real time while digital signals are sent in bursts of pulses, stored, and released to recreate the original sound. The horizontal blanking interval of the video signal transmits the digital audio channels and one or more utility data channels. The digital audio bit stream is encrypted according to an appropriate algorithm adding to a key. Downlinks possessing the correct key translate the binary digits into the original signal.

Through a process known as line translation, segments of each digitized line of video are sampled by the encoder and converted into digital values. The digitized line segments are cut and rotated so that the segments within each line are shuffled out of order and reassembled at either side of the cut points. Each line has different cut points; all vertical information in the picture is broken up: stepped back and forth across the screen with each line and in a sequence that changes from field to field. A Pseudo Random Binary Sequence generator that is synchronous with both the uplink transmitter encoder and downlink receiver decoder identifies the cut points for each line. As the uplink encoder periodically interrupts and restarts the final control algorithm, the encoder must send a special seed code to the decoder to indicate re-initialization. This keeps the decoder dynamically locked to the uplink Pseudo Random Binary Sequencer. New seeds are generated and transmitted periodically to each authorized decoder in the system. This seed, which also is encrypted, is transmitted within the television signal's vertical blanking interval [Anderson, 1997]. The algorithms needed to unlock the encrypted seeds are either embedded in a tamper-proof medium within each decoder or supplied by the previously mentioned smart security card.

From the previous description of the satellite encryption process for video and audio, it is necessary to mention that the transfer of digital data may undergo similar processing without the need for the format conversion of analog-to-digital prior to transmission. Therefore, the bandwidth requirements to support the all-digital data stream may be much less than those required to support video and / or audio transmissions.

**Summary**

Historically, the focus of network specialists has been primarily upon the terrestrial assets when regarding system security issues. However, it must be noted that the major network in use currently also entails the use of certain extraterrestrial assets with expansion of that capability planned for the immediate future. Therefore, the identification of satellite security accommodations and a firm understanding of the relevant technologies are necessary in order to complete an end-to-end networking security plan for organizations with a global interest.

Signal security is the culmination of all aspects of the transmitted intelligence that mitigates or decreases the risk of interception and divulgence. It is assumed in this document that environmental influences upon the transmitted signal, such as space related ultraviolet ray harmonic distortion of signals and atmospheric influences as the Aurora Borealis, are negated. Additionally, influences related to orbital mechanics are also assumed to be null in this research.

**Significant Contribution**

This survey attempts to encompass progressive technologies and methods in an effort to offer options to minimize response / recovery costs, operational costs, and maximize the supportability to the integrated response team of government and volunteer support personnel. Additionally, the data system must assure the continued capability while phasing from the recovery support and returning to the legacy, reinstated infrastructure asset capability. The portability of the recovery system and the use of national infrastructure capacities that remain available near the disaster area to form an integrated, seamless data distribution network during contingency efforts offers lower operational costs and efficient introduction of transmitted data into the remaining fully capable infrastructure network.

This document draws upon insight into systems presently utilized by the National Communication System to support other governmental agency requirements. Most of the satellite technologies characterized in this document are under the jurisdiction of the National Aeronautics and Space Administration, Goddard Space Flight Center. Systems that were examined included the Tracking and Data Relay Satellite (TDRS), DomSat, and the Advanced Communications Technologies Satellite with the encryption techniques described reflective of non-military accommodations (National Aeronautics and Space Administration, 1999). Research of end-to-end satellite data transfer was also made possible through the National Aeronautics and Space Administration, John F. Kennedy Space Center and Space Transportation System (Space Shuttle) mission STS-95.

This research was undertaken through the examination of governmental reports of actual cases of satellite communications to satisfy a requirement for data transfer from remote sites or areas where public terrestrial infrastructure service was unavailable. Research into specific encryption techniques was also undertaken in an effort to best present the system capabilities for security during satellite communications. Discussion of one widely used technology for encryption of uplink information was provided with convolutional encoding, whereas, the downlink discussion was rendered as exemplified in the discussion of Viterbi decoding.

A primary scientific precept is that in an effort to provide a complete understanding of the technologies involved, one must understand the nature of the subject. A satellite, as related to this analysis, is a space borne platform used as a remote relay for information or to act as a station for remote sensing of terrestrial events and features. Currently in the general technology, satellite systems support the broadcast of intelligence via radio or subscriber services to remote or mobile receivers as described in articles on the XM and Sirius technologies, as well as television and data transmission relay (Lewis, 2000). However, the security of these services continues to be of paramount importance in current society but none as important as during any crisis or emergency action. Satellite data may come in the form of services from supporting terrestrial communication by telephone, fax, or computer, observation of the world's weather, and remote sensing for use by researchers and cartographers to make more accurate maps. Therefore, data integrity must be assured as to allow the data to be used in decision-making processes with confidence.

When a satellite is launched, it is placed in orbit around the Earth using the Earth's gravity to hold the satellite in the required path. Three prominent types of orbits will be discussed here, in order, Low Earth Orbit (LEO), Medium Earth Orbit, and Geosynchronous Earth Orbit (GEO) (American Telephone &Telegraph, 2000). Due to the speed of the spacecraft while in orbit, most communications occur in a burst mode of transmission thereby offering a narrow window of opportunity for interception and decryption. This characteristic of data transfer also provides an element of security when the initial data packet received includes crucial decryption information thereby offering interception and decryption unlikely in most satellite communications.

Security is provided through a myriad of encoding and compression algorithms during the transmit sequence for both analog and digital operations thereby offering rendering the integrity of the transmitted intelligence.

Given the diversity of encryption schemes that are based upon those previously mentioned in this research, the potential for unauthorized decryption is unlikely for most attempting access. The computing power and the costs associated that must be assumed during an attack of satellite data would require resources generally outside of those of common system intruders.

Additionally, the highly specific equipment required to intercept satellite downlink streams are both costly and require infrastructure support that would preclude a stealthy intrusion into the data stream or downlink antenna beam footprint.

It is the findings of this research that the nature of the transmission precludes most intrusion and thereby infers a level of security in the transmission that is not offered through the terrestrial infrastructure. However, this research does not address the use of the satellite communications network to perpetuate an attack upon a terrestrial network. This is due primarily to the designed nature of the satellite as a relay device as opposed to a space bourn server or data interrogation device.

Although much of the operation of the satellite systems rely upon ground system commanding, no specific mention is made of the techniques in place for assuring the integrity of uplinked commanding and downlink of telemetry. This is predominantly due to the sensitive nature of this topic and the small amount of data publicly available. Many of the techniques mentioned earlier may be applied to the control of spacecraft systems in varying degrees, however, these techniques are often applied in tandem with other more specific algorithms.

Security of satellite systems continues to be of concern to organizations that service and are supported by this technology. Current technologies available to the public that support untethered business processes including personal digital assistants and cellular telephones and offer a new basis for business founded upon current information. However, as society continues to rely upon near real-time intelligence in order to remain competitive in the global marketplace, the need for alternative technologies capable of supporting during contingency periods remains. Furthermore, the need to provide security for information regardless of the transfer device becomes increasingly important.

Encryption standards and the costs associated with any attempt to breach satellite communications makes the probability of attack unlikely. However, as technologies evolve in the current trend, the potential for more portable and powerful transmission/reception and decryption mechanisms increases daily. Alterations of various parameters of the data stream makes the present potential low, but how long will this Utopia continue?

As the technology advances, the IT professional must continue the awareness of the potential for security breach and remain proactive in the search for the proper combination of tools to meet the impending threats. Vigilance is the watchword in today's computer based economy. Care and planning must become the key for every system operator and administrator-improved planning and education of users, as well as, the network technical staff is imperative for the survivability of the network, the information contained, and perhaps, the business that it serves.

It is imperative that research continues in the vein to improve cryptography to support the extraterrestrial networking. Additional work is also required in the arena of intra-satellite communications using laser technologies and other optical techniques due to the potential for speed and resistance to the effects of the spatial environment. Additionally, progress must be made in the capability to assure autonomous satellite operation thereby precluding the need for ground control intervention. This capability would then preclude the tampering of satellite service availability from terrestrial hacking or sabotage.

**Acknowledgements**